\documentstyle[namedreferences,psfig]{spacekap}
\begin{opening}
\title{On the difference between radio loud and radio quiet AGN}
\author{Karl Mannheim}
\institute{Universit\"ats-Sternwarte,
Geismarlandstr. 11, D -- 37083 G\"ottingen, Germany\\
E-mail: kmannhe@medusa.uni-sw.gwdg.de}
\date{}
\end{opening}
\begin{document}
\begin{abstract}
Nuclear jets containing relativistic ``hot" particles
close to the central
engine cool dramatically by producing high
energy radiation.  The radiative dissipation is similar
to the famous Compton drag acting upon
``cold'' thermal particles in a relativistic bulk flow.
Highly relativistic
protons induce anisotropic showers raining electromagnetic power
{\it down} onto the putative accretion disk.
Thus, the radiative signature of hot hadronic jets
is x-ray irradiation of cold thermal matter.  The synchrotron
radio emission of the accelerated electrons is self-absorbed due to the
strong magnetic fields close to the magnetic nozzle.
\end{abstract}
\section{Jets and accretion disks:  Castor and Pollux?}
A puzzling mystery for AGN theorists is the relation
between the big blue
bump emission component, which is believed to originate
as thermal emission
from matter surrounding a supermassive black hole
and the emission related to the morphological appearance
of jets.
Recent $\gamma$-ray
observations have shown that
the blazar spectrum from the subparsec jet
is quite different from a thermal one and
extends over
almost twenty orders of magnitude in frequency with
an almost constant
level $\nu S_\nu$.
Amazingly, the properties of the big blue bump alone
never show any indication
of whether or not the AGN also has a powerful jet.
On the other hand,
BL~Lacs show no sign of a big blue bump at all.
On the basis of these facts one is tempted to assume that
there are two hearts in AGN:  one beating for the
thermal processes
(high entropy) and one for the nonthermal processes
(low entropy).
But then
we remember our aim as physicists is to simplify and not
to secularize the
physical world.  Could there be a relation between the
two phenomena, in the
sense that one is, perhaps, more fundamental than the other?
This
question has been around for some time during which the paradigm
changed from the
nonthermal origin of activity to the thermal one,
the accretion paradigm.
If the latter is correct, then
viscosity must be totally robust with respect to the generation
of a powerful jet.

In this context
I find two new results most
challenging to our understanding of AGN.  Firstly,
$\gamma$-ray
measurements teach us that the prime radiation
mechanism in jets
seems to favour extremely high energies:
Mkn421 was observed at an
energy of 1~TeV with a
$\nu S_\nu\propto \rm const.$ spectrum.
Secondly, a fairly robust relation $Q_{\rm j}=100
L_{\rm nlr}\approx L_{\rm bb}$ between the
kinetic power of jets
$Q_{\rm j}$
and the narrow line luminosity, resp. the photoionizing
luminosity of the big blue bump, in radio galaxies
and quasars has been found by
\citeauthor{Rawlings91} \shortcite{Rawlings91}
and \citeauthor{Celotti93} \shortcite{Celotti93}.
They tell us
that whatever links nonthermal and thermal
components does this
independent of the total luminosity and the presence of broad lines or a
big blue bump.
The existence of pure objects of either class then
appears inconsistent.
In fact, radio quiets do have weak radio jets and a nonthermal
optical/x-ray
continuum.
So, one should perhaps take a different
point of view:
assuming {\it all}
AGN have powerful jets
with
$Q_{\rm j}\approx L_{\rm bb}$,
could
their radiative
appearence be different enough to account for all AGN classes
\cite{Camenzind83}?
In
particular, this requires that the jets in radio quiets must
dissipate
most of their kinetic power witin the central parsec, because
further out
they would unescapably show strong radio emission.
The momentum, however, is
still there and, indeed, radio quiets do show high
speed nuclear outflows,
most spectacularly in broad absorption line quasars.
{}From this heuristic starting point the following questions
arise:
(i)  if jets in radio loud sources generally start out
with relativistic speeds (which can plausibly be assumed with
$\gamma_{\rm j}\le 10$), is
this a prerequisite for the jets to emerge out of the
central parsec without suffering much from radiative losses?
If so,
(ii)  what is it that lets some jets start out slower,
(iii)  why do slower jets dissipate so much of
their power in the
central parsec and
(iv)  why is this connected to the properties
of the host galaxy?

In this contribution I will attack question (iii), while
leaving the answers to the other questions open.
My personal guess would be that question (i) must be answered
affirmatively, probably because relativistic bulk motion
protects the jet from shock acceleration
too close to
the disk.  Concerning (ii)
it is quite natural for
any jet forming mechanism to generate something like
an inverse power law distribution for the speeds of jets,
so that one easily gets 90\% mildly relativistic jets
and 10\% relativistic jets.  And finally, (iv) may have
to be reversed in order:  different jet properties make
different host galaxies because jets can trigger
star formation via cosmic rays and shocks.  Remnant winds
in quiets would sweep out interstellar matter above the disk
leaving behind a torus.
\section{Beamed $\gamma$-rays:  a fingerprint from accelerated protons}
In an effort to empirically answer the question whether jets contain
an ordinary mixture of protons and electrons, \citeauthor{Mannheim91}
\shortcite{Mannheim91}
investigated the radiative signature of highly relativistic protons
accelerated at shock fronts.
Photoproduction of pairs
and pions
injects electromagnetic power into the
acceleration zone which is further reprocessed by
an (unsaturated) synchrotron cascade.  This proton initiated
cascade (PIC) should operate at shocks of all sizes:  starting
from kpc Hot Spots (cf. Harris {\it et al.}, this volume) down to the shocks
at the subparsec scale thought to be responsible for ``proton blazar"
emission
\cite{Mannheim93}.

An essential parameter is the distance of the proton blazar
from the source of the big blue bump photons, because local
photons compete with thermal photons from outside the jet as a
target.  As the proton acceleration zone moves closer in, the
$\gamma$-ray spectrum steepens as shown in Fig.(1) until cooling is
entirely dominated by the anisotropic thermal target photons.
\begin{figure}
\centerline{\psfig{figure=mannheim.fig1.ps,height=5cm,width=13.7cm}}
\caption{The proton blazar model for 3C273 with a proton/electron ratio
of $\eta=15$, see Mannheim, {\it Phys. Rev. D}, Vol.48, No.4 (1993).
Note the steepening of the $\gamma$-ray
spectrum due to the additional blue bump target photons.}
\end{figure}
\section{Dying jets:  hadronic shower precipitation and
nuclear winds}
Protons exposed to an anisotropic target field cool via photoproduction
mostly in the direction of the source of the target photons.  They
prefer head-on collisions with $\mu=-1$ because of the
threshold condition
$\gamma_{\rm p}x(1-\beta_{\rm p}\mu)\ge x_{\rm k,th}'$
where $\mu=\cos\theta$ is the cosine of the angle between
proton and target photon momentum, $x=h\nu/m_{\rm e}c^2$
the photon energy and $x_{\rm k,th}'$
the photon threshold energy  in
the proton rest frame
to create particle k$=e^\pm,\pi$ on the mass shell.
Thus, head-on collisions require the
lowest $\gamma_{\rm p}$ and thus produce the strongest flux for
a proton distribution $n_{\rm p}\propto \gamma_{\rm p}^{-2}$
cooling on an almost monoenergetic photon target $x\approx 2
\times 10^{-4}$.  The required proton Lorentz factor is
$7\cdot 10^5$ for pion production.  Figs.(2) shows the
emergence of this {\it natural anisotropy}.
Maximum efficiency of the
irradiation is obtained at the distance to the disk
of $z_\circ= 200\sqrt{r_\circ/100}$
(in units of the Schwarzschild radius)
where the photoproduction optical depth becomes unity for a jet radius
of $r_\circ$, but $\tau_{\rm pp}\ll 1$.
At this location the magnetic field
has the strength $B_\circ=1.6\cdot 10^3M_{\rm A}^{-1}
m_8^{-1/2}r_\circ^{-1}\beta_{\rm 0.3}^{-1/2}$ and the target
radiation compactness is $l_\circ=3m_8^{-1}r_\circ^{-1/2}$.
Here it was assumed that
$L_{\rm edd}=L_{\rm bb}+Q_{\rm j}+
L_{\rm B}+L_{\rm rel}\simeq L_{\rm bb}+L_{\rm B}\left(M_{\rm A}^2
+2\right)$
with equipartition $L_{\rm B}=
\left(B^2/8\pi\right)A_{\rm j}c\beta_{\rm j}=L_{\rm rel}$
and the kinetic power $Q_{\rm j}=
M_{\rm A}^2L_B$.
The irradiation spectrum is a powerlaw with $\alpha\simeq 1$
from eV up to GeV (because of a lack of x-ray photons to reprocess
$\gamma$-rays from MeV to GeV) and
$L_{\rm pic}=L_{\rm bb}(1+M_{\rm A}^{-2})/
\left(0.5+M_{\rm A}^2\right)$.  A mildly relativistic jet
suffering from such severe radiative losses rapidly expands.
The surviving momentum is shared to the surrounding thermal
matter driving a high speed nuclear
wind.
Radio emission from the accelerated electrons is
synchrotron-self-absorbed up to the frequency
$\nu_{\rm s}=4\cdot 10^{13}m_8^{-1/3}r_\circ^{-1}M_{\rm A}^{-4/3}
\beta_{0.3}^{-2/3}$ which makes such dying jets radio quiet.
There is some radio emission associated with the remnant wind
and the low emissivity reflects its large opening angle.
\begin{figure}
\centerline{\psfig{figure=mannheim.fig2.ps,height=5cm,width=13.7cm}}
\caption{{\bf Left panel:}  The angular distribution of cascade power.  Note
the
transition from emission into the forward Doppler cone $\mu=\beta$
to emission
into the backwards hemisphere when the bulk Lorentz factor $\gamma$
decreases.  At very small angles no emission is produced because there
are no photons satisfying the
threshold condition.
{\bf Right panel:}
The backward/total luminosity ratio as a function of jet
Lorentz factor.  A maximum value of $\approx 90\%$ irradiation is possible.
The label $k=4$ denotes the curve for a proton maximum energy four times
the threshold energy for head-on collisions.  Thus, for $\gamma>4$ no
photoproduction ocurrs.  Close to the limit the total luminosity is small,
but emitted solely towards the  backward hemishere.
}
\end{figure}
\section{Critique and conclusions}
In this contribution I have proposed that powerful jets
are ubiquitous in AGN.  Radio quiets are equipped with
mildly relativistic jets which cannot emerge from the
central parsec because of severe radiative losses and rapid expansion.
Relativistic
protons in these jets induce electromagnetic
showers which irradiate the accretion disk with hard radiation and
which have very weak emission towards the observer.
An {\it experimentum crucis} is the detection of the neutrino
signature of the hadronic processes
which is feasible with
experiments currently under construction \cite{Stenger92}.
The model needs a high efficiency of converting kinetic energy
into radiation which is difficult to reconcile with statistical
acceleration mechanisms (Mastichiadis \& Kirk, this volume).  Magnetic
dissipation could resolve the problem, but could
it yield proton Lorentz factors of $10^6$?

\end{document}